# Using a Neural Network to Detect Anomalies given an N-gram Profile

Byunggu Yu and Junwhan Kim

Computer Science and Information Technology
University of the District of Columbia, Washington DC 20008, USA
{byu, junwhan.kim}@udc.edu

**Abstract.** In order to detect unknown intrusions and runtime errors of computer programs, the cyber-security community has developed various detection techniques. Anomaly detection is an approach that is designed to profile the normal runtime behavior of computer programs in order to detect intrusions and errors as anomalous deviations from the observed normal. However, normal but unobserved behavior can trigger false positives. This limitation has significantly decreased the practical viability of anomaly detection techniques. Reported approaches to this limitation span a simple alert threshold definition to distribution models for approximating all normal behavior based on the limited observation. However, each assumption or approximation poses the potential for even greater false positive rates. This paper presents our study on how to explain the presence of anomalies using a neural network, particularly Long Short-Term Memory, independent of actual data distributions. We present and compare three anomaly detection models, and report on our experience running different types of attacks on an Apache Hypertext Transfer Protocol server. We performed a comparative study, focusing on each model's ability to detect the onset of each attack while avoiding false positives resulting from unknown normal behavior. Our best-performing model detected the true onset of every attack with zero false positives.

## 1 Introduction

Over the last four decades, the software assurance community has developed various types of monitoring techniques in order to detect intrusions and errors in computer programs at runtime. Such monitoring techniques focus on the following approaches: (1) detecting known bad patterns, (2) detecting runtime deviations from the design specifications of the program, or (3) detecting runtime deviations (anomalies) from the observed normal of the program.

The first approach (known bad pattern detection) [2, 8, 22, 27, 31, 34] encompasses predominantly signature-based matching employed by various anti-virus and network packet monitoring solutions. Attack signatures are usually built by analyzing known attacks. The process of collecting signatures implicitly assumes and accepts the risk that any unknown attack has to succeed or become detectable at least once. Only after the attack can any relevant information be collected, studied, and shaped into a signature to be used in the detection. Sometimes, the initial signature needs to be improved later to increase the effectiveness of the signature in detecting variants of the corresponding attack or to address any logic that was not initially discovered by the analysts. The signature-based approach does not generally detect or protect against exploits of zero-day vulnerabilities unless they happen to match already known malicious signatures. The second approach, design specification-deviation detection [1, 24, 40] requires a priori knowledge about the target program's internal logic and additional efforts to generate corresponding behavior specifications.

Unlike the first two approaches, the third approach, anomaly detection [5, 13, 15] does not require a priori knowledge of attack behavior or application design in order to protect an arbitrary computer program from unknown intrusions and errors. The basic rationale is as follows: if the entropy of the behavior of the program is finite, one can observe the program to learn about its normal behavior and then detect anomalies that do not follow this behavior. Unfortunately, there are three practical challenges with this anomaly detection approach.

1. The first challenge is **the representation of normal behavior**. There have been investigations of using the names of system calls generated by a program to represent a trace of a program's behavior, e.g. [15]. However, it has also been reported that mimicry attacks [16, 25, 41] can go undetected by mimicking some normally observed sequence of system calls. A feature space is a set of attributes of a system whose values are being observed. If the granularity of the feature space is insufficient, it may be impossible to distinguish between the normal and abnormal, lowering the true positive rate.
2. The second challenge is that **practical observations are often limited** and may not cover all possible normal behaviors of the target program. This practical limitation often results in false positives during detection, when alerts are issued for the normal behaviors that were not previously observed. Excessive false positive alerts can make the detection solution unusable. On the other hand, any generalization of the learned normal can create false negatives regarding anomalous behaviors covered by the generalization of the normal behavior, lowering the true positive rate.
3. The third challenge is to **learn normal behavior using noisy behavior observations** in an unsupervised manner. Any observed noise can incur false negatives regarding any intrusions or errors that resemble the noise. Such false negatives will correspondingly lower the true positive rate [10].

This paper focuses on anomaly detection methods that address these three challenges. The proposed detection methods take a bit stream as input, where each bit has the value **0** if the corresponding local behavior matches a known behavior or **1** otherwise. Although a variety of models for behavior representation can be considered, the system call **n**-gram model [15] is used in this paper. Due to the factors mentioned in the challenges described above, the resulting **n**-gram match-mismatch bit stream is inherently noisy. Therefore, the main problem investigated in this paper is how to reliably find anomalies in this noisy bit stream.

We present and compare four methods that are designed to detect anomalies in an **n-**gram match-mismatch bit stream. We report here on our experience running various types of attacks several times on a relatively lightly loaded Apache Hypertext Transfer Protocol (HTTP) server application. We processed the server's system call behavior, before, during and after the attacks, using the nine models. Then we performed a comparative analysis, focusing on each model's ability to detect the attacks while avoiding false positive alarms.

The results presented in this paper show that our proposed method, called LAF (denoting "LSTM Anomaly Filter"), is highly practical in terms of false positives, true positives, detection delays, supported types of programs, and detectable types of attacks. By eliminating false positives without compromising the true positive rate, the LAF can detect the true onset of DoS attacks early (e.g., almost immediate detection attacks with no false positives in our experiments). This makes timely remedial actions possible, with enough time and computational power. Moreover, other types of attacks involving a smaller number of system calls can be addressed.

The rest of this paper is organized as follows: Section 2 defines the problem; Sections 3 presents the four anomaly detection methods; Section 4 presents our experimental setting and evaluates our results; Section 5 summarizes related work. Then, Section 6 concludes the paper with a summary and our vision of future work.

## 2    Problem Definition

System call (syscall) anomaly detection approaches tend to include two conceptual phases:
- **Training**: creates a behavior model (or profile) for a sequence of consecutive syscalls made by a process during its normal states. A controlled environment is often used to assure normal operation of the process.
- **Detection**: an anomaly alert is issued when the observed sequence of syscalls deviates from the behavior model by more than an allowed threshold. This paper is focused on the quantification of the model-appropriate threshold selection.

We collect and record the n-grams of the syscall streams for each program to be monitored. During the monitoring phase, runtime syscall n-grams undergo a match test over the recorded n-grams and a match (0)/mismatch (1) bit stream is produced. In determining the normality of the runtime behavior, we apply a distribution-based classifier approach.

Let $\hat{p}$ be the average match rate of **n-**grams at which the training is arbitrarily set to complete (i.e., defining the end of the training phase).

Let $X \sim \mathbf{Binomial}(W, \hat{p})$ be a random variable modeling the number of matches (i.e., zeros in the bit stream) in $W$ consecutive **n-**gram match tests. If the "aggregation window size" $W$ is large enough and $\hat{p}$ is not too close to either **1** or **0**, the binomial distribution can be considered approximately equivalent to a Gaussian distribution with the following mean and variance:

$$\mathbf{E(X) = a = W\hat{p}} \tag{2.0.1}$$

$$V(X) = \text{var}(X) = E\left((X - E(X))^2\right) = \sigma^2 = W\hat{p}(1 - \hat{p}) \qquad (2.0.2)$$

Under the assumptions the relation or random variable and the distributions can be presented as:

$$X \sim \text{Binomial}(W, \hat{p}) \approx \mathcal{N}(W\hat{p}, W\hat{p}(1 - \hat{p}))$$

This expression means that random variable **X** has binomial distribution, which in turn is approximated by a normal distribution with expected value **W$\hat{p}$** and variance **W$\hat{p}$(1 − $\hat{p}$).** For this approximately Gaussian distribution, it generally makes sense to consider that a system is in a normal state when the number **X** of matches in the binary coded match/mismatch string of length **W** is within range **E(X) ± m$\sigma$** of the mean of the corresponding normal distribution. However, in our application, a match rate above upper bound **E(X) + m$\sigma$** is also considered to be normal because observing signs of normal behavior in abundance is normal. Hence, we'll focus on the match rates falling below the lower bound **E(X) − m$\sigma$.** We will use **3$\sigma$** anomaly threshold, hence take fixed **m=3**.

Let **n** be the length of the **n-grams**. A completely new **n-gram** in the worst case can ultimately result in up to **2n-1** consecutive mismatches if it consists exclusively of previously non-profiled syscalls. Therefore, the size W of the aggregation window should be large enough to accommodate at least **2n-1** consecutive mismatches in a normal state. That is, W should be set such that:

$$W - (W\hat{p} - m\sqrt{W\hat{p}(1 - \hat{p})}) \geq 2n - 1. \qquad (2.0.3)$$

Here W is the aggregation window size, **n** is the length of each **n-gram**, and **m=3, for example,** is the multiplier allowing anomaly detection with the **3$\sigma$** span.

Based on this basic definition of the problem of anomaly detection, the paper presents three anomaly detection methods given – EWMA (Exponentially Weighted Moving Average Method), PEWMA (Probabilistic EWMA), and LAF (LSTM Anomaly Filter).

## 3   Anomaly Detection Methods

3.1  Static Binomial Method (SB)

For the first model we make the simplest assumptions about the bit stream. Any isolated bit stream can be naïvely viewed as a realization of some Bernoulli test sequence with a "success" (or match in our case) probability. For a large enough sequence of experiments Bernoulli distribution can be approximated by a normal distribution with parameters given by **(2.0.1)** and **(2.0.2)**. The training stage provides us a reasonable estimate of the match probability **$\hat{p}$**, which we assume is good enough to characterize the process in a normal state. So, to build the simplest, while admittedly naïve model for bit stream anomaly detection, we assume that the process and observable bit stream data possess the following three properties:

(1) The probability of **X** is approximated by the normal distribution;

(2) The individual match/mismatch tests are described by the Bernoulli distribution;
(3) The match probability $\hat{p}$ is static (i.e. does not change over time).

If these assumptions hold, we can consider the process model to be in a normal state if the following holds:

$$X \geq W\hat{p} - m\sqrt{W\hat{p}(1-\hat{p})} \tag{3.1.1}$$

**W** is set such that, the inequality **(2.0.3)** is satisfied.

On one hand the simplicity of the expression **(3.1.1)** allows for easy computation of an anomaly condition. On another hand the assumptions (1)-(3) that lead to the expression may be too naïve as independence of the consequential **n**-grams is an intentional simplification for the Bernoulli model's necessary conditions, so SB is excluded in our experiment.

3.2. EWMA

In Section 2 we assumed that the process behavior was static, i.e. distribution of the matches describing normal process behavior was specific only to the process and did not change with time. In our dynamic models, we will not seek or assume existence of one distribution universally describing normal behavior of the process.

One way of addressing assumption (3) from the **SB** is to reevaluate the probability mass function of **X** sequentially as new observation data comes in. However, simply substituting $\hat{p}$ of the previous models with localized estimates may misclassify the corresponding local anomalies. Hence, the history of **X** needs to be included into localized statistics. The localized statistics introduced in the sections below can allow us to estimate distribution parameters specifying normal process behavior from one point in time to another.

To allow localized computing of an average value, we will consider EWMA (the Exponentially Weighted Moving Average [33]), which is a popular localized averaging technique that computes the local average $E(X_t)$ for a time point t by recursively applying exponentially decreasing weights to the past averages as follows:

$$E(X_t) = \alpha E(X_{t-1}) + (1-\alpha)X_t, \tag{3.2.1}$$

where $0 \leq \alpha < 1$ is the weight put on the history. Based on the normal distribution assumption, $X_t$ is considered to meet the following condition in most cases:

$$|E(X_t) - X_t| < m\hat{\sigma}_t, \tag{3.2.2}$$

where **m** is a constant multiplier (as before we use $3\sigma$) and $\hat{\sigma}_t$ is the localized estimate of standard deviation computed for a point **t** in time as follows:

$$\hat{\sigma}_t = \sqrt{E(X_{t-1}^2) - E(X_{t-1})^2} \tag{3.2.3}$$

In our application, $X_t$ is defined to be the number of matches during **W** consecutive match/mismatch tests. Therefore, $E(X_t) < X_t$ is not considered to be anomalous, but rather **normal above average** behavior. Based on this observation and **(3.2.2)**, we can consider the process to be in a normal state if the following holds:

$$X_t \geq E(X_t) - m\hat{\sigma}_t, \qquad (3.2.4)$$

where, $E(X_t)$ is computed according to **(3.2.1)**, $\hat{\sigma}_t$ is computed according to **(3.2.3)**.

3.3. PEWMA

It has been reported that EWMA tends to be optimized at a higher end of the $\alpha$ value range (i.e., close to **1**) in terms of mean squared error (MSE) prediction given a data stream containing a small number of anomaly events placed near each other with short inter-occurrence times [4]. Under such an optimization, the mean **E(X)** changed by a large anomaly does not return to the normal level fast enough to detect closely following smaller anomaly events. To address this [4] proposed a variant of EWMA called Probabilistic EWMA (PEWMA). This model replaces the localized mean recursive update expression **(3.2.1)** as follows:

$$E(X_t) = \alpha(1 - \beta P_t)E(X_{t-1}) + (1 - \alpha(1 - \beta P_t))X_t, \qquad (3.3.1)$$

where $0 < \alpha < 1$ is the history weighting parameter, $P_t$ is the probability of $X_t$ under some modeled distribution (as in [4] we use a standard normal distribution), and **β** is the weight placed on $P_t$.

It is easy to see that with $\beta \to 0$ or $P_t \to 0$ the expression for PEWMA converges to EWMA. The rationale behind PEWMA is that samples that are less likely to have been observed should have lesser influence on the corresponding updates. In order to accomplish this based on the normal distribution, $P_t$ is defined as follows: $P_t = \min\{(1/\sqrt{2\pi})e^{(-\frac{(E(X_{t-1})-X_t)^2/\hat{\sigma}_t^2}{2})}, \hat{p}\}$, where $\hat{\sigma}_t$ is computed as in **(3.2.3)** and $\hat{p}$ is the expected match rate recorded at the completion of the training.

The process is in a normal state if the following holds: $X_t \geq E(X_t) - m\hat{\sigma}_t$. Even though this expression appears to be the same as that of EWMA, here the localized mean is defined as **(3.3.1)**. As before **m=3** is used as the constant multiplier for anomaly detection.

3.4. LSTM Anomaly Filter (LAF)

The most annoying problem in any threshold-based anomaly detection mechanism is caused by recurrent match-mismatch patterns that can trigger false positive alerts. The problem becomes more cumbersome as the number and complexity of such patterns increase. Of course, this problem diminishes when our n-gram profile is complete and covers 100% of all normal behavior of the target program. However, in theory, this is hard to justify. Given the fact that we do not have the source code of the program, we cannot assume the language type the program's system call sequences. If this language belongs to, say, context-free language type, the n-gram model, which is a sub-regular expression, will not be able to represent all system call sequences in a finite set of n-grams.

In order to address recurring false positives incurred by recurring match-mismatch sequence patterns, we need an anomaly detection technique beyond and above the statistical threshold approaches. In this case, we need one that has a learning capacity: the recurring false positives are quickly learned and have the associated recurring alerts diminish quickly over time. More learning capacity is required as the number, pattern complexity, and the occurrence complexity of the false positive match-mismatch sequences increase.

Long and Short-Term Memory (LSTM) as a special recurrent neural network (RNN) with an input, hidden and output layers has been utilized for long-range dependencies [47]. On the hidden layer of LSTM at time, the outputs – $c^{t-1,l}$ and $h^{t-1,l}$ of the previous layer at $t-1$ come in the layer at $t$ as inputs. The major advantage of LSTM controls a cell status $c^{t-1,l}$ that indicates an accumulated state information. The cell state is updated or cleared by several operations. If this state is cleared, the past cell status is forgotten by $f^{t,l}$. If updated, $c^{t,l}$ – one of the outputs at $t$ will be propagated to the final state. The cell state is prevented from vanishing or exploding gradient, which is a problem of the traditional RNN, resulting in more learning capacity. Like PEWMA, LAF starts with match-mismatch sequences with a training phase.

## 4    Application Experiment

Given a bit stream of system call (syscall) **n-**gram match test results, each model presented in Sections 2 through 3 defines a condition that the bit stream must satisfy in order to qualify as normal. For the purpose of anomaly detection, the system is instructed to issue an anomaly alert whenever the model-specific conditions are not satisfied. For quantitative verification and validation, we implemented the models and installed the implementation on an operational Apache HTTP Server environment. The syscall **n-**gram model [9, 14, 15, 35, 37, 42, 45, 46] was used for the training and match tests with **n=6** [37].

In this application test, we upgraded the original syscall **n-**gram technique to support multi-process and multi-threaded applications. The Apache HTTP Server uses multiple process instances at runtime. To accommodate the multi-process and multi-threaded nature of the application, the **n-**gram training was modified to aggregate the syscalls into process and thread groups to assure that each syscall **n-**gram belongs entirely to a particular execution context. This modification significantly improved the training time as arbitrary interleaving between threads were not learned nor matched against.

During the training phase, the model implementations observed the web server during normal operation periods and profiled every unique **n-**gram. The training phase was set to finish when there are fewer than **23** mismatches over the last 10000 **n-**grams, which would yield p̂ corresponding to **0.9973**. The W value (aggregation window size) of the simple models (EWMA and PEWMA) varied between 1700 and 5000. The $W$ value of the stream models varied between 20 and 5000. Note that the minimum W value is limited in the simpler models (e.g., SB) as given in their corresponding models. This is not the case in the stream-based models. This is one of the advantages of using stream-based models when it comes to detecting short attacks. We revisit this in the Conclusions section.

Based on reported test cases [4], the α values of EWMA and PEWMA were set to **0.97**, 0.99, respectively, giving heavy weight on recent history to detect anomaly onsets. In order to conduct live attacks[1], we used an application-level, remote denial-of-service (DoS) attack identified by CVE-2011-3192, to which the target web server was vulnerable. The tested implementation of this attack (available at http://www.exploit-db.com/exploits/17696/) accepts an input parameter called *numforks*, which defines the intensity of the attack. By exploiting the HTTP protocol, this application-layer DoS attack causes memory exhaustion within the application. In our test, a high-intensity attack was remotely initiated and immediately terminated. The second attack was performed after some period of time at a lower intensity level and lasted until the system became unresponsive.

4.1. Results from Aggregations

In our tested cases, this value was in the 20-100 range. Figure 1 shows our experimental results with EWMA, PEWMA, and LAF assigned as label *a*, *b*, and *c*, respectively. EWMA and PEWMA can be applied with aggregation windows. This is advantageous for two reasons: (1) the inherent delay between the actual onset point of an anomaly and the corresponding alert point can be reduced; (2) has the potential to detect different attack types involving a small sequence of system calls. Our aggregation tests shown in Figure 1 revealed that the proposed LAF is the best performer with zero false positives. However, it was found that LAF's attack detection delay tends to increase as the aggregation size increases.

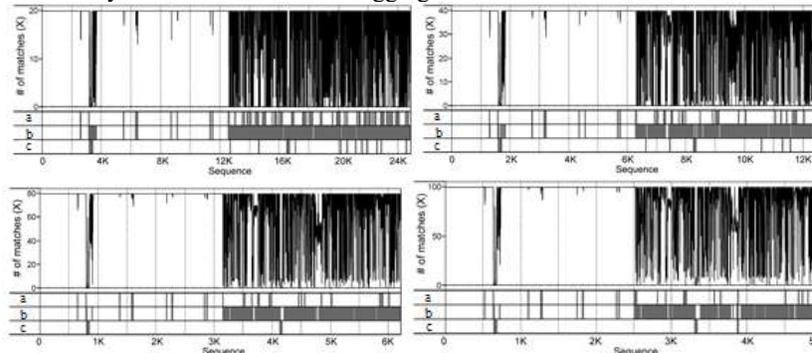

Figure 1. Anomaly Detection Models with Aggregations: N=20, 40, 80, and 100

4.2. Slow HTTP

Most of our empirical work measures the behavior of "fork bombs" in which crafted input can cause a victim web server to exhaust local process tables. There are, however, a variety of strategies for denying service. To gain generality in our results, in this section we measure the effectiveness of our detection models when they are subjected to "slow HTTP" attacks.

Slow HTTP attacks (https://github.com/shekyan/slowhttptest/wiki) exploit a natural asymmetry between HTTP clients and servers: a client can send requests to an HTTP server in an incremental way so that, at low cost to the client, a server must allocate and maintain significant system resources such as socket send buffers. Slow HTTP attacks are significant because the vulnerability arises from resource policies implemented by numerous HTTP servers (e.g., keeping very low-flow connections alive) for legitimate quality-of-service reasons. As a complement to our "fork-bomb" experiments for evaluating our detection models, slow HTTP attacks are illuminating because their implementing mechanisms are entirely different.

---

[1] The experiments were conducted with the permission of the system owner.

We tested the effectiveness of three of our models (EWMA, PEWMA, and LAF) by running slow HTTP attacks and measuring each model's false positives and negatives with different training periods. We chose these models for the slow HTTP tests because these models work well with small aggregation windows. For each model, we ran three tests with varying levels of training performed prior to each test. At the high (expensive) end, our training phases continued until the cumulative match rate reached .9973 (very little behavior not seen during training). At the low end, our training only reached a very incomplete cumulative match rate of .8.

In Figure 2, we show the results from our three models assuming the highest level of training (.9973 cumulative match rate). The five lines at the top of the figure show the aggregation window size (5) and show graphically the number of mismatches (up to the size of the aggregation window) over time (left to right). The bottom three columns of the figure show the alerting behavior of the three models: EWMA, PEWMA, and LAF. As can be seen, an early false positive on the left hand is generated by the first two models but not by LAF. At this level of training, all three models detect the onset of the attack (but not quite at the same time).

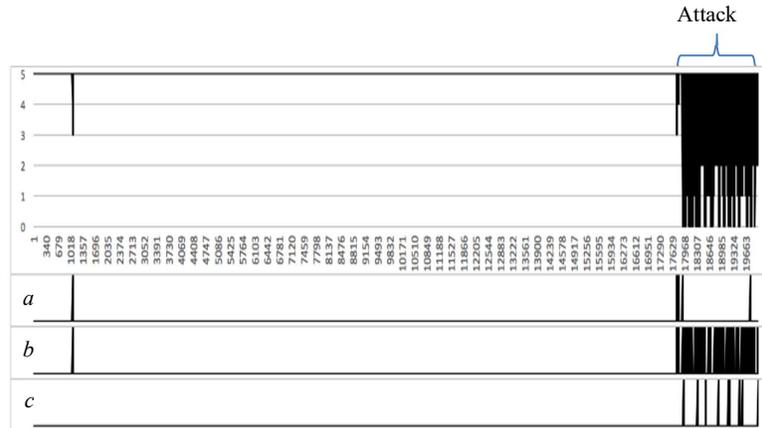

Figure 2. Slow HTTP attack with 99.73% training

In our second test configuration (Figure 3), we measure detection model performance when the training is completed with a cumulative match rate of .9545. With less complete training, one can see that there are numerous mismatches over the period of observation. The first two models (EWMA, PEWMA) generate a large number of false positives in addition to detecting the onset of the actual attack, however LAF generates no false positives and also identifies the onset of the attack.

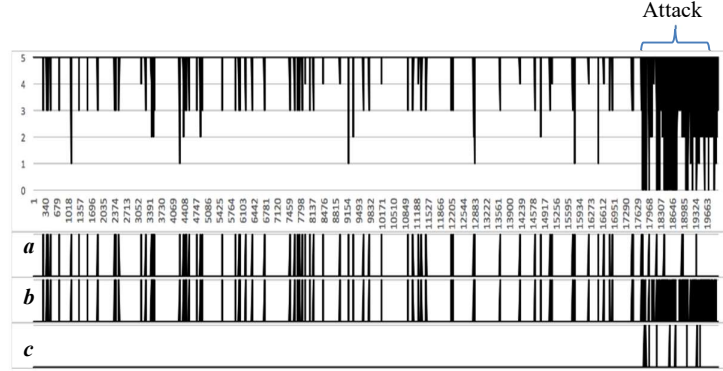

Figure 3. Slow HTTP attack with 95.45% training

In our third test configuration (Figure 4), we measure detection model performance when the training is completed with a cumulative match rate of only .8. In this case, the number of mismatches is overwhelming as one would expect with such incomplete training. The first two models (EWMA, PEWMA) generate a continuous stream of false positives. The LAF model, however, generates no false positives and also detects the onset of the attack with only a minor delay. This is a striking outcome: how can this model be so much more effective?

We believe that this is due to the fact that a neural network addresses the overlapping and probabilistically dependent nature of the **n**-grams on a more fundamental stochastic level than distribution-based models. An **n**-gram following a normal **n**-gram is more likely to be normal. Conversely, an abnormal **n**-gram is more likely to be followed by an abnormal **n**-gram [29]. In our match-mismatch bit stream representation, this means that a mismatch is more likely to be followed by a mismatch, rather than a match. Therefore, contrary to the assumptions (1) and (2) made in Section 2 for SB, the bit stream is not very accurately represented by the Bernoulli distribution due to the interdependence between the **n**-grams. Meanwhile, the neural network provides an evaluation model of the process' stochastic behavior independent of distribution. This is the most prominent distinction of the neural network, when compared to the traditional distribution-based classification methods [18].

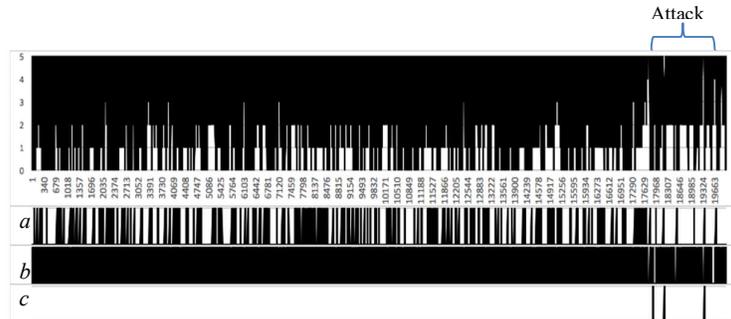

Figure 4. Slow HTTP attack with 80% training

## 5    Related Work

We have studied various anomaly detection methods for profiling normal behavior in terms of system calls during the training-phase and for using the recorded profile during the detection-phase to confirm the normality of incoming runtime sequences. Some reported methods of system call (syscall) anomaly detection are based on **n-**gram and statistical learning models [15, 42]. Other reported methods are based on automata models and call-stack return-address models [13, 35, 40]. Ideally, the anomaly detection methods should be able to complement existing signature or misuse detection systems to enhance the overall security effectiveness of the system by detecting unknown errors and attacks for which signatures are not yet defined.

Unfortunately, application behavior data observed during the training phase is often limited to a small subset of all normal behavior or possibly tainted by hidden attacks or errors [10, 11, 18, 21]. Therefore, it is virtually impossible to prove a claim that the behavior model captures all necessary aspects of the behavior in such a way that all possible unknown attacks and errors can subsequently be distinguished. This paper organizes reported approaches to this challenge into two groups: (1) classifier group and (2) sensitivity group.

The approaches from the classifier group generalize the limited training data to certain distributions or feature space regions in order to more uniformly classify the unknown behaviors. Related studies include statistical generalization and classification of the distributions of the data by normal distribution [4], machine learning [10], kNN methods [10, 30, 26], and support vector machines [7, 11, 18].

The approaches from the sensitivity group adjust the anomaly alert threshold in order to correctly ignore benign noise in the detection phase. This group includes stide, t-stide, and the specific use cases of RIPPER and HMM reported in [42]. The sequence time-delay embedding (stide) method takes a sequence of syscalls as input and stores the sequence from the training-phase as unique **n-**grams with most frequently selected value of **n** set to 6. In the detection phase, each **n-**gram, representing a sliding window of the runtime syscall sequence, is compared with the training phase collected **n-**gram sequences. The anomaly alert threshold is defined in terms of the number of mismatches in the "locality frame" (the reported case is the last 20 match tests). An anomaly alert is issued only when the runtime sequence incurs mismatches above the threshold. This method is most effective when clean, normal behavioral sequences are used for training. The threshold sequence time-delay embedding (t-stide) method is a variant of stide, which drops rarely occurring **n-**grams from the trained normal. The reported example is dropping any **n-**gram that accounts for less than 0.001% of the total number of **n-**gram occurrences counted during the training [23, 24]. In [42], t-stide was less effective compared to stide. However, it was shown in [10] that t-stide is more effective when trained over noisy data.

A specific use case of RIPPER in which a mismatch is defined to be a violation of a high-confidence rule (e.g., violation score greater than 80 [28] is also reported in [42]. The anomaly alert threshold is defined in terms of the number of mismatches over the last 20 syscalls. In the Hidden Markov Model (HMM), the alert threshold is defined in terms of a minimum required probability associated with transitions and outputs [42].

Reported findings [15] show the following: when the threshold is set high enough, (1) the average false positive (FP) rate of the **n-**gram models is similar to the RIPPER rule-learning model, and much better (lower) than that of a HMM; (2) the true positive (TP) rate of **n-**gram models is much better compared to RIPPER although worse than HMM. The overall performance characteristic of the **n-**gram models is that both TP and FP rates decrease as the threshold increases (i.e., TP becomes worse while FP becomes better).

By storing and matching the **n-**grams of a runtime sequence without significant computation, **n-**gram models naturally pose modest computational overhead compared to statistical learning models. However, the anomaly alert threshold, with its potentially adverse effect on TP and FP rates, is the key performance parameter.

## 6   Conclusion

The experimental evidence supports the idea that LAF is desirable solutions in terms of the Receiver Operating Characteristics (ROC) of anomaly detection in the presented application. In other words, the proposed solutions showed a minimized false positive rate without compromising the true positive rate (maximum true positive rate and minimum false positive rate). The large aggregation of the match test bit stream can reduce the computation overhead of the anomaly detection. However, this approach comes with the inherent delay in anomaly detection in terms of the number of system calls (syscalls) between the true onset of an anomaly and the detection incurred by the aggregation over time. Moreover, different types of attacks, such as shell-code attacks, involving a small number of syscalls can be washed out by the surrounding normal behavior in large aggregations. Because of this, we believe that a method supporting rather small aggregation is likely to be more useful in practice.

The proposed LAF is the best performing detection model based on our experimental data and analysis. Our application and experimental results have shown that LAF is a highly practical anomaly detection solution in terms of ROC, detection delay, and supported types of attack. By supporting any aggregation size with effective elimination of false positives without compromising the true positive rate, LAF can detect the onset of DoS attacks early with minimal false positives (e.g., almost immediate detection of DoS attacks with no false positives in our **W**=20 cases). This makes remedial action possible, with enough time and computational power. Moreover, other types of attacks involving a smaller number of syscalls can be addressed. This case is outside the scope of the paper and represents future work.

### Acknowledgements

The authors acknowledge and appreciate Dmitry A Cousin, David Waltermire, and Lee Badger at the National Institute of Standards and Technology (NIST) for their review and comments on an early development version of this article. The authors also appreciate Dong H. Jeong regarding his help in the visualization of some of our experimental results.